\documentclass[%
 reprint,
 amsmath,amssymb,
 aps, superscriptaddress,
]{revtex4-1}
\usepackage{xcolor}
\usepackage{graphicx}
\usepackage{dcolumn}
\usepackage{bm}
\usepackage{setspace}
\usepackage{subcaption}

\begin{document}

\title{Spatial control of extreme ultraviolet light with opto-optical phase modulation}

\author{\mbox{A. Olofsson$^{1}$}, \mbox{E. R. Simpson$^{1}$}, \mbox{N. Ibrakovic$^1$}, \mbox{S. Bengtsson$^1$}, \mbox{J. Mauritsson}}

\address{Department of Physics, Lund University, P.O. Box 118, SE-221 00 Lund, Sweden.}

\begin{abstract}
Extreme-ultraviolet (XUV) light is notoriously difficult to control due to its strong interaction cross-section with media. We demonstrate a method to overcome this problem by using Opto-Optical Modulation guided by a geometrical model to shape XUV light. A bell-shaped infrared light pulse is shown to imprint a trace of its intensity profile onto the XUV light in the far-field, such that a change in the intensity profile of the infrared pulse leads to a change in the shape of the far-field XUV light. The geometrical model assists the user in predicting the effect of a specific intensity profile of the infrared pulse, thus enabling a deterministic process.

\end{abstract}

\maketitle

Manipulating light by inserting optical components that change the wavefront is perhaps the oldest form of conscious manipulation of light. This has, in the modern age, developed into the invention of optical modulators such as acousto-optic modulators and electro-optic modulators, where the refractive index of a medium traversed by the light can be modified with great precision. This permits a high degree of control over the wavefront of the light. The resulting precision has lead to numerous possibilities of shaping visible light; amplitude modulation, phase modulation, spatial modulation, frequency modulation and polarization modulation \cite{SalehPhotonics2007}. However, these techniques, and any others that are based on common refractive optics, are not suitable for extreme-ultraviolet (XUV) light due to the strong interaction cross-section in this range. The very reason why XUV light absorbs so strongly in media is the same reason that it is so useful as a spectroscopic tool for atomic and molecular studies: it spans the energy range permitting single photon excitation or ionization for a wide range of electronic states, including many situated in inner electronic shells \cite{Attwood_Soft_X_Rays_and_Extreme_Ultraviolet_Radiation}. It is therefore of importance to investigate techniques that could bridge the difficulty in controlling XUV light.

Control of XUV light requires a way to modify the phase of the light such that the desired shaping occurs. This has been demonstrated through control over the density profile of an interacting gas \cite{DrescherNature2018}, effectively changing the optical path-length of different spatial regions of the XUV light. The technique to control XUV light that we focus on here, however, is opto-optical modulation (OOM), where a non-resonant light pulse is instead used to control the XUV in an interaction gas \cite{Bengtsson2017, BengtssonJ.Phys.B.2019}. The reversal has also been shown, where the behaviour of the XUV light is used as a probe into the interaction between the gas and the non-resonant pulse \cite{SimpsonPhysRevA2019}.

The OOM technique controls light in the far-field by manipulating the phase of the light in the near-field, similar to many optical components that are based on Fourier optics. For XUV light, it is not suitable to use for example a phase-shifting mask based on transmission \cite{Mack_FieldGuideToOpticalLitography} but instead, OOM uses near-field modulation that occurs through an XUV pulse resonantly exciting an ensemble of atoms followed by a non-resonant infrared (IR) control pulse, which alters the phase of the excited ensemble. At the moment of excitation by the XUV pulse, the ensemble of excited atoms is coherently forced into oscillatory motion as the superposition between high-energy states and the ground state evolves in time \cite{Bengtsson2017, BrewerPhysRevA1972, Scully.PhysRevA.2105}. The temporal evolution of the phase of the ensemble depends on the energy separation between the ground state and the excited states, while the absolute phase depends on the phase of the incident XUV pulse. The atoms then undergo coherent, radiative decay.

After a variable time delay, a non-resonant control pulse is directed onto the atoms. Due to the AC Stark effect, the control pulse shifts the energy separation between the ground state and the excited states \cite{StarkNature1913}.  When the energy separation is changed, this change is transferred to the absolute value of the phase. The change in the phase of the emitted light is given by:
\begin{equation}
\Delta\phi(\mathbf{r}_a) = \frac{1}{\hbar}\int_\tau \Delta E(\mathbf{r}_a,t)dt
    \label{Eq:PhaseShift}
\end{equation}
where $\Delta\phi(\mathbf{r}_a)$ is the phase shift for a specific atomic state at the specific point $\mathbf{r}_a$ where the atom is located, $\tau$ is the duration of the interaction between the atom and the control pulse, and $\Delta E(\mathbf{r}_a,t)$ is the instantaneous energy shift of the specific atomic state. $\Delta E(\mathbf{r}_a,t)$ depends on the intensity of the control pulse and the polarizability of the atomic states involved in the superposition \cite{DelonePhysUsp1999}. This means that the resulting phase is a convolution between the intensity profile of the control pulse and the phase response of the Stark-shifted state. In this article we are focusing on a few closely spaced high-lying $p$-states in Helium, from 1\textit{s}8\textit{p} at 24.38 eV \cite{Hopfield1930_7p}. Previous studies show that in general, high-lying $p$-states behave approximately ponderomotively, that is their phase shift is approximately linear with the intensity of the control pulse \cite{SimpsonPhysRevA2019}. Henceforth we will confine the discussion to the behaviour of these high-lying \textit{p}-states in Helium and approximate their phase shifts to be equal to the ponderomotive phase shift, given by the energy shift $\Delta E(\mathbf{r}_a,t)\propto I(\mathbf{r}_a,t)$ \cite{BucksbaumJOptSocAmB1987}.

\begin{figure}[tbh]
\begin{centering}
\includegraphics[width=0.3\textwidth]{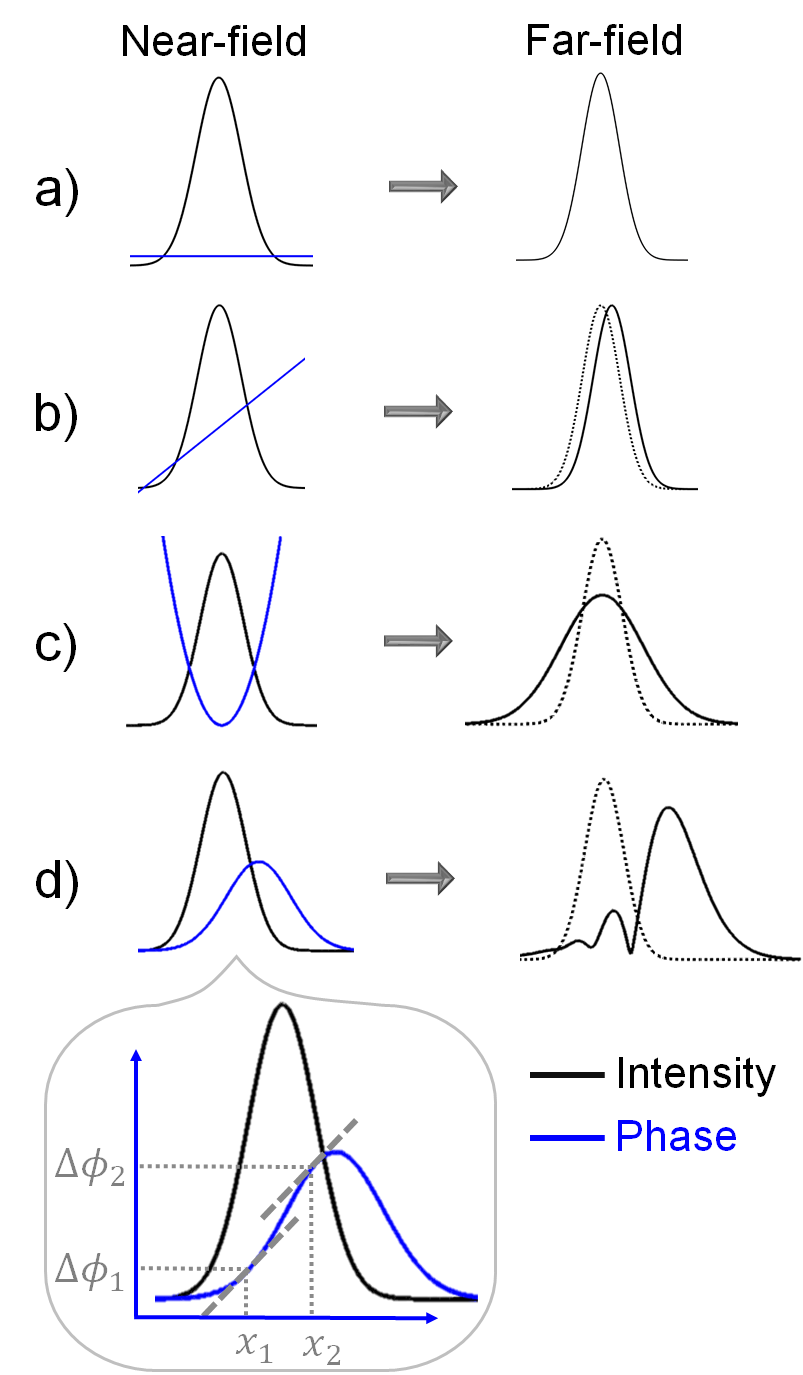}
\caption{Examples of far-field shaping of an initially Gaussian pulse through near-field phase modulation. In b)-d), the dotted line shows the initial Gaussian centered around $x=0$ for clarity. For a Gaussian phase profile in the near-field, such as in d), the resulting intensity profile in the far-field shows interference fringes. This interference results from equal gradients at different locations of the near-field profile, as marked in the inset, with different magnitude of the applied phase. Note that the phase magnitude and the magnitude of the intensity profile are scale-adjusted for clarity. 
}
\label{fig:Shaping}
\end{centering}
\end{figure}

To see what this means in practice, it is instructive to first look at a few examples. We begin with the simplest scenario - an XUV pulse with a Gaussian spatial intensity profile and a constant phase applied across. If a constant phase shift is applied to the Gaussian XUV in the near-field (see Fig.~\ref{fig:Shaping}a), the XUV intensity profile in the far-field will be a Gaussian that is identical to the case of no applied phase shift. However, if the applied phase shift varies over the spatial extent of the near-field Gaussian XUV intensity profile, the resulting far-field intensity profile will be modified. For example, a linear gradient across the near-field profile gives a spatially displaced intensity distribution in the far-field (see Fig.~\ref{fig:Shaping}b). A quadratic phase shift applied in the near-field can be seen to result in a changed width of the far-field XUV intensity profile (see Fig.~\ref{fig:Shaping}c). In the experimental setup, the intensity profile of the IR control pulse is approximately bell-shaped and therefore it is of special interest to discuss what happens if the applied phase is bell-shaped. This is explored in Fig.~\ref{fig:Shaping}d: in this example both the initial XUV intensity profile and the applied phase have Gaussian spatial profiles. The applied phase has a slightly broader full-width at half-maximum (FWHM) than the intensity profile does, as we would expect in the experimental setup. The resulting far-field XUV intensity profile is displaced and shows interference fringes. The reason for the interference is that it is the gradient of the resulting XUV wavefront that controls the direction of propagation, so if the applied phase is shaped in such a way that it imprints the same gradient for two spatial locations, different parts of the near-field intensity profile will propagate to the same spot in the far-field. Two locations with the same gradient but with different absolute phase shifts will lead to interference in the far-field and fringes can be observed.

\begin{figure}[h]
\includegraphics[width=0.51\textwidth]{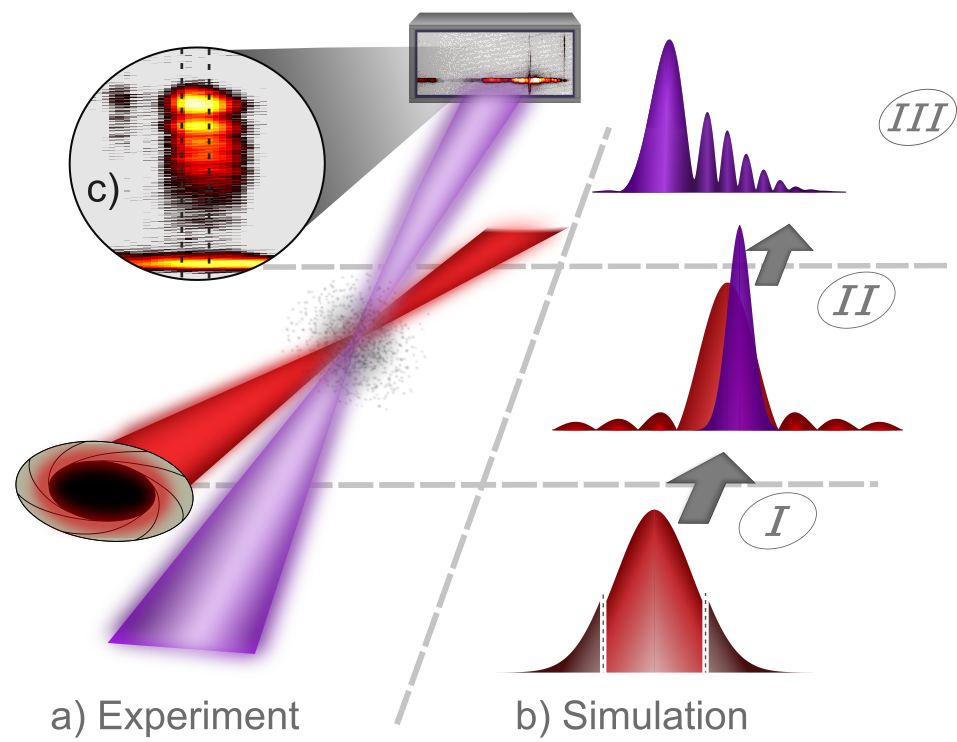}
\caption{Schematic presentation of the experimental process (a) and the simulated steps (b) in the geometrical model. Row I shows the variable cutting of the IR control pulse (red). The arrows in (b) represents far-field transforms. Row II describes the imprinting of the phase shift onto the XUV pulse (purple). Row III shows the detector plane where the spatial XUV intensity profile is recorded for each aperture opening. Finally, the dotted lines in the inset (c) show the frequency region that is integrated from the experimental data to produce the plot in Fig.~\ref{fig:ExpVsTheory}a. Note that the angle between the XUV beam and the control beam in (a) is exaggerated for clarity, and the grating in the experiment is neglected.}
\label{fig:ExpVsSim}
\end{figure}

Experiments were undertaken using the setup described in \cite{Bengtsson2017, BengtssonJ.Phys.B.2019, SimpsonPhysRevA2019}. The control pulse is an IR pulse that is non-concentric and slightly non-colinear with the XUV light, but overlaps spatially with the XUV light in the mutual focus. The spatial offset between the centers of the two pulses in the focus was optimized for maximum redirection of the XUV light at the detector. The delay of the IR control pulse was set to arrive just after the XUV pulse to allow a strong redirected signal without significant IR dressing during the XUV excitation. The part of the experimental setup that is of particular interest to this research begins with the IR control pulse with approximately bell-shaped spatial intensity profile being cut by a variable aperture, see a(I) of Fig.~\ref{fig:ExpVsSim}. In the model, this is simulated by symmetrically truncating a Gaussian IR intensity profile around the center, see b(I) in Fig.~\ref{fig:ExpVsSim}. 

\begin{figure}[h]
\includegraphics[width=0.5\textwidth]{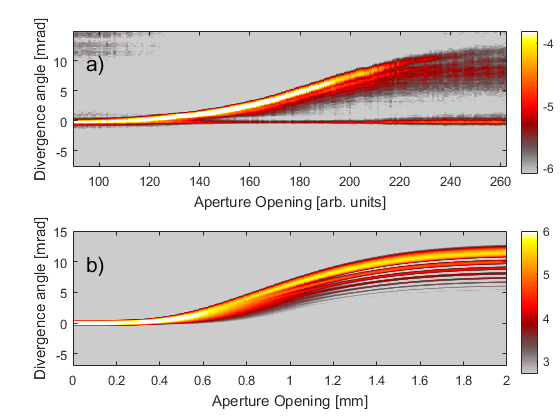}
\caption{Experimental results (a) from scanning an aperture opening of an IR control pulse imprinting a nonlinear phase shift onto an atomic ensemble emitting XUV radiation. The data for each column has been divided by the sum of all signal in the column to compensate for fluctuations. The result can be reproduced theoretically by using a geometrical model (b) described in the text. Note that for the experimental case, the opening of the aperture is controlled by rotation, meaning that the increase in aperture diameter is not necessarily linear as is the case for the model. For both plots the intensity is depicted in logarithmic scale.}
\label{fig:ExpVsTheory}
\end{figure}

The next step in the process is the focusing of the XUV pulse (with approximately bell-shaped intensity profile) with a platinum-plated toroidal mirror into a pulsed valve distributing Helium in a low-pressure environment. This is where the OOM process takes place. The IR control pulse is focused into the Helium by the same toroidal mirror as the XUV. In the model this step is simulated by propagating the control pulse that is transmitted through the aperture with a far-field transform, based on the Fraunhofer approximation, to the focus. The focused IR control pulse intensity profile is used to calculate the accumulated phase shift from the Stark effect, under the assumption of ponderomotive behaviour and a temporally Gaussian intensity profile. This accumulated phase shift is then imprinted on a spatially Gaussian XUV intensity profile. Another far-field transform is then used to take this modified XUV intensity profile to the detector plane, row III in Fig.~\ref{fig:ExpVsSim}. In the experiment, the modified XUV pulse propagates to a flat-field reflective grating and is then detected by micro-channel plates, a fluorescent screen and a camera. In the data analysis, a narrow frequency range is chosen (as shown in Fig.~\ref{fig:ExpVsSim}c) and the selected region is integrated to show the signal for each divergence angle for each aperture opening, which gives results such as those presented in Fig. \ref{fig:ExpVsTheory}a. Note one difference between the experiment and the model in that in the model, only one state and one frequency is simulated, which corresponds to only one point on the frequency-separated axis of the experimental results. However, in the experimental results the different high-lying \textit{p}-states are not resolved and instead form a frequency band.

The data resulting from scanning an aperture from closed to open are shown in Fig.~\ref{fig:ExpVsTheory}a. The figure shows a frequency range (marked by dotted lines in Fig.~\ref{fig:ExpVsSim}c) from the camera corresponding to the emission from high lying $p$-states in Helium, and the horizontal axis denotes the opening angle of the motorized aperture that the IR control pulse passes through. It should be noted that the sensitivity of the detector is reduced at an emission angle around $0^{\circ}$. 
Although there are differences between the model and the experiment, we find that the simple geometrical model can assist in predicting how the XUV light will be controlled by the IR.
Figure~\ref{fig:ExpVsTheory}b is produced using spatially Gaussian profiles for both the XUV pulse in the focus and the control pulse (before the aperture), with the XUV FWHM in the focus set to 50 $\mu$m, and the spatial offset to the control pulse in the focus also set to 50 $\mu$m. The FWHM of the IR pulse in the focus, for a fully open aperture, is set to 120 $\mu$m, and the temporal FWHM of the control pulse is set to 25 fs. To match the experimental data, the peak intensity of the control pulse is set to $5.1\cdot 10^{13}$ Wcm$^{-2}$ in the focus (for the case of a fully open aperture), which gives a maximum accumulated phase difference of 128 radians. The mask that simulates the aperture opening is varied from a diameter of 0 to 2.3 mm. It should be noted that while the size and shape of the pulses in the simulation are set to simulate the experiment as well as possible, the FWHM:s and shapes in the focus are not exactly known in the experiment, and the intensities of the pulses are also not exactly known. The maximum intensity value used in the simulation was chosen to match the maximum gradient to that of the experimental data, which is detected through the maximum divergence angle. The on-axis emission present in the experimental results in Fig.~\ref{fig:ExpVsTheory}a also after $x\approx 140$  can be accounted for due to signal measured at the detector from atoms that did not interact with the control pulse.  This on-axis signal is not reproduced in the theoretical data presented in Fig.~\ref{fig:ExpVsTheory}b as these non-interacting atoms are not taken into account in the simulation.
The experimental data is likely also affected by signal dampening at high intensities due to for example ionization of the gas, which is not taken into account in the simulation.  
Despite these differences, the simple geometrical model has proven to be a useful and intuitive tool to understand the experimental results. Without the complexities of more complete theoretical calculations, the simple geometrical model provides a fast understanding of the experimental results.

The model confirms the hypothesis from Fig.~\ref{fig:Shaping} that if the XUV intensity profile overlaps spatially with the control pulse intensity profile in the focus such that multiple points in the overlap share the same control pulse intensity gradient, there will be interference effects in the XUV intensity profile at the detector as seen in Fig.~\ref{fig:ExpVsTheory}. When the aperture opening is small, the peak intensity and the gradient of the IR control pulse will be low. This results in a smaller redirection from the on-axis light. Further, when the aperture size is small, the FWHM of the IR in the focus is large and thus the interference effect diminishes in this case. 
 When the aperture opening increases, the maximum intensity gradient increases and redirects the modulated XUV emission further and further away from the on-axis emission. At the same time, the spatial extent of the IR control pulse in the focus decreases and the overlap region can cover more and more of both the highly nonlinear region toward the center of the bell-shaped control pulse intensity profile and the fringe of the intensity profile. This evolution with aperture opening can be seen in Fig.~\ref{fig:ExpVsTheory}, for both the experimental results a) and the simulated results b). This is a clear sign of how the spatial intensity profile of the XUV pulse in the far-field is shaped by the spatial intensity profile of the IR control pulse in the near-field. A remark can also be made for the potential use of the fringes seen at x $\approx$ 250 in Fig.~\ref{fig:ExpVsTheory}a: since each full fringe marks when the phase difference between the up-turned part of the bell-shaped control pulse and the down-turned part of the same has changed by $2\pi$, it might be possible to use the fringes in combination with calculations of the intensity-dependence of the Stark shift to determine the peak intensity of the control pulse in the focus.

To summarize this work, we have taken a step further on the road to full control of XUV light through OOM, from purely redirecting a chosen temporal part of the XUV emission to now being able to use the spatial intensity profile of the IR control pulse to deterministically shape the XUV intensity profile in the far-field, with the model as a guide. A natural way forward would be to experiment with different intensity profiles of the control pulse, for example using a spatial light modulator, to extend the XUV-shaping possibilities.

We would like to acknowledge the importance of the phase shift calculations \cite{SimpsonPhysRevA2019} performed by Seth Camp, Marie Labeye, Kenneth Schafer and Mette Gaarde at Louisiana State University.

This research was supported by the Swedish Research Council, the Crafoord Foundation, the Birgit and Hellmuth Hertz Foundation/The Royal Physiographic Society of Lund, and the Wallenberg Center for Quantum Technology (WACQT) funded by The Knut and Alice Wallenberg Foundation (KAW 2017.0449).

\bibliography{ABibliography.bib}

\pagebreak

\end{document}